\documentclass[%
 reprint,
 amsmath,amssymb,
 aps,
floatfix,
]{revtex4-2}

\usepackage{graphicx}
\usepackage{dcolumn}
\usepackage{bm}
\usepackage{mhchem}
\usepackage{float}
\usepackage{hyperref}

\begin{document}

\preprint{APS/123-QED}

\title{Conditional Generative Models Enable Targeted Exploration of MAX Phase Design Space}

\author{Jamie Swaine}
\affiliation{%
Department of Chemistry, 
University College London, 
Kathleen Lonsdale Building, 
Gower Pl, London, WC1E 6BS, UK.
}

\author{Cyprien Bone}
\affiliation{%
Department of Chemistry, 
University College London, 
Kathleen Lonsdale Building, 
Gower Pl, London, WC1E 6BS, UK.
}

\author{Matthew T. Darby }
\affiliation{%
AWE Aldermaston, Reading, Berkshire, RG7 4PR, UK. 
}

\author{Ewan Galloway }
\affiliation{%
AWE Aldermaston, Reading, Berkshire, RG7 4PR, UK. 
}

\author{Keith T. Butler}
\email{k.t.butler@ucl.ac.uk}
\affiliation{%
Department of Chemistry, 
University College London, 
Kathleen Lonsdale Building, 
Gower Pl, London, WC1E 6BS, UK.
}

\date{\today}

\begin{abstract}
MAX phases (M$_{n+1}$AX$_n$), precursors to MXenes, span a vast compositional space, motivating efficient computational screening for synthesisable candidates. We employ CrystaLLM$-\pi$, a large language model fine-tuned on 6,179 double transition-metal MAX phases, and demonstrate its ability to generate out-of-sample structures consistent with known experimental trends. Using a conditioning vector with two dimensions (a statistically derived MXene derivative count and a surrogate for A-site binding energy), the model was able to target MXene-favourable regions of phase space for generation. Specific condition vectors double novel stable structure generation rates versus unconditioned baselines. Of ten compositionally novel candidates, five exhibit DFT-validated stability ($E_{hull} < 0.050$ eV/atom).  This work showcases the potential for autoregressive generative models to explore targeted materials' spaces, offering a scalable framework for accelerated discovery in compositionally complex systems.
\end{abstract}

\maketitle

\section{Introduction}

MAX phases constitute a class of nano-laminated materials that combine ceramic-like robustness with metallic conductivity. They have found use in many applications, including corrosion- and radiation-resistant coatings for nuclear materials,\cite{lambrinou2017} energy storage,\cite{alam2024} and catalysis.\cite{chirica2021} Represented by the general formula M$_{n+1}$AX$_n$, MAX phases comprise an early transition metal M-site, an A-site which is generally a group 13-14 element, and a non-metal X-site, which is most often carbon, nitrogen, or boron. Scientific interest in these phases has increased rapidly in recent years, with 342 experimentally synthesised compositions reported as of 2024, approximately half of which were discovered after 2018. Notably, MAX phases also serve as precursors to MXenes, two-dimensional derivatives that retain many of the parent MAX characteristics while offering tunable surface chemistries. Represented by the general formula M$_{n+1}$X$_n$T$_x$, MXenes arise from the selective removal of the A-site element and the subsequent formation of surface terminations, T, whose identity depends on the synthetic and post-processing environments.

As of 2024, 28 M-site elements, 28 A-site elements, and 6 possible X-site elements have been reported in experimentally synthesised MAX phases.\cite{dahlqvist2024} Even under a single stoichiometric family the naïve expansion of the element sets leads to more than 4,700 theoretical ternary compounds before accounting for variation in the crystal structure itself. Consequently, it is both efficient and necessary to identify and evaluate theoretically stable candidates via computational methods, thereby refining and guiding experimental search. 

The application of machine learning techniques in chemistry has increased sharply over the past decade, with such methods now frequently integrated with traditional theoretical and computational approaches to alleviate the computational bottleneck associated with \textit{ab initio} calculations \cite{batatia2023, zeni2025generative, ong2019accelerating, merchant2023scaling, walker2026carbon}. Deep neural networks have demonstrated accuracy comparable to that of density functional theory (DFT) while performing evaluations orders of magnitude faster.\cite{li2022, batatia2025design} In particular generative models offer the promise of predicting new viable materials, with desired properties, on demand \cite{sanchez2018inverse}.

Recently there have been a number of studies reporting transformer-based "large language models" (LLMs) adapted for materials design \cite{antunes2024, gruver2024fine, flam2023language}. These approaches leverage the flexibility and scalability of text-based transformers for modelling complex chemical space.\cite{mohantyCrysTextGenerativeAI2025, breuckGenerativeMaterialTransformer2025, kazeev2503wyckoff, alampara2024mattext}

Beyond transformer-based approaches, a range of contemporary frameworks for crystal structure generation have emerged. Physics-guided deep learning models that incorporate explicit symmetry and stability constraints have been of particular interest.\cite{millerFlowMMGeneratingMaterials2024, pakornchoteDiffusionProbabilisticModels2023, jiaoCrystalStructurePrediction2024, jiaoSpaceGroupConstrained2024, klipfel_diff_aaai_2024, levySymmCDSymmetryPreservingCrystal2025, cornetKineticLangevinDiffusion2025, chang2025space, nguyen2023hierarchical, zhao2023, parkExplorationCrystalChemical2024} These models offer stronger physical grounding and reduce the need for extensive post-processing. Together, generative and physics informed approaches demonstrate a growing shift toward computational strategies that tightly couple machine learning with underlying materials chemistry and physics.

In this study, we consider the application of CrystaLLM,\cite{antunes2024} a transformer-based autoregressive generative model, trained on Crystallographic Information Files (CIFs) which encode lattice parameters, space-group, and atomic species. This model was recently updated to include conditioning mechanisms as of 2025 allowing for property guided generation.\cite{bone2025discoveryrecoverycrystallinematerials} This updated conditioning model is subsequently referred to as CrystaLLM$-\pi$. Under conditional generation this model produces crystal structures in the same way as the base model, but with the additional objective of satisfying user specified targets, encoded in the condition vector. Here we condition CrystaLLM$-\pi$ to explore specific regions of MAX phase design space.

A secondary investigation was also conducted on MAB phases, an under-explored subclass of MAX phases that have recently attracted attention for similar applications to those described previously, but possess boron in the X site. This represented a low-shot learning regime, within which CrystaLLM$-\pi$ successfully generated novel structural motifs that were likewise validated through DFT calculations at the PBE level.

As part of these investigations, we utilise the GNNs (Graph Neural Networks), MACE (Message Passing Atomic Cluster Expansion) and ALIGNN (Atomistic Line Graph Neural Network) for energy and property predictions. MACE and ALIGNN are members of a class of models known as MLIPs (Machine Learned Inter Atomic Potentials), which have been shown to offer accuracy on the order of DFT calculations at a cost similar to deriving classical interatomic potentials.\cite{Leimeroth2025MSMSE}

The results of this study demonstrate how the capabilities of CrystaLLM$-\pi$ can be leveraged for targeted exploration of MAX phase chemistry. Across tens of thousands of structural generations, we show that model conditioning not only promotes broader exploration of the compositional and structural design space, but also yields crystal structures that exhibit strong fidelity to their imposed conditioning. Furthermore, CrystaLLM$-\pi$ was shown to generate structures with out-of-sample chemical composition that were also thermodynamically stable, under DFT at the PBE level of theory. 
The results highlight the utility of conditioned autoregressive transformer-based models for materials design and discovery.

\section{Results and Discussion}

Two conditioning properties were employed in this study. The first served as a preliminary screening step, assessing structures based on their potential MXene derivative count. The second condition involved a sublattice perturbation of A-site elements, used as a surrogate for vacancy formation energy and free-energy evaluation.

A MAX phase was designated as having a derivative by cross-referencing the fine-tuning dataset with the MXene dataset from aNANt. Two key assumptions were made in calculating the derivative count for any given species. First, the stoichiometry of the parent compound was not assumed to be preserved during etching.\cite{benchakar2020} Second, the nature of MXene terminations was assumed to be dictated entirely by the etching and post-processing environment, and thus fully customisable.\cite{bao2021} On this basis, a MXene derivative was defined as any subset of the transition-metal elements comprising the parent MAX phase.

Based on these assumptions, 6,013 structures from the fine-tuning dataset (not including the MAX/MAB-like structures) were identified as possessing between 1 and 172 potential MXene derivatives. This condition vector is entirely statistically derived and simply serves to guide generation with respect to elemental contents of the MAX phase. This count was normalised using a min-max scaler between 0 and 1, meaning that a value of 0.00 would aim to push generation towards a space associated with the lowest number of derivatives, where a value of 1.00 would suggest the greatest derivative count. 

To derive a surrogate for the binding energy of the A-site, and thus obtain a meaningful representation of vacancy formation and free-energy values, a perturbation-evaluation procedure was applied to the A-sublattice. The A-site layer was displaced by 1\% of the $c$ lattice parameter in either direction, and for each perturbation the $E_{hull}$ metric was computed using MACE.

\begin{figure}[h!]
    \centering
    \includegraphics[width=\columnwidth]{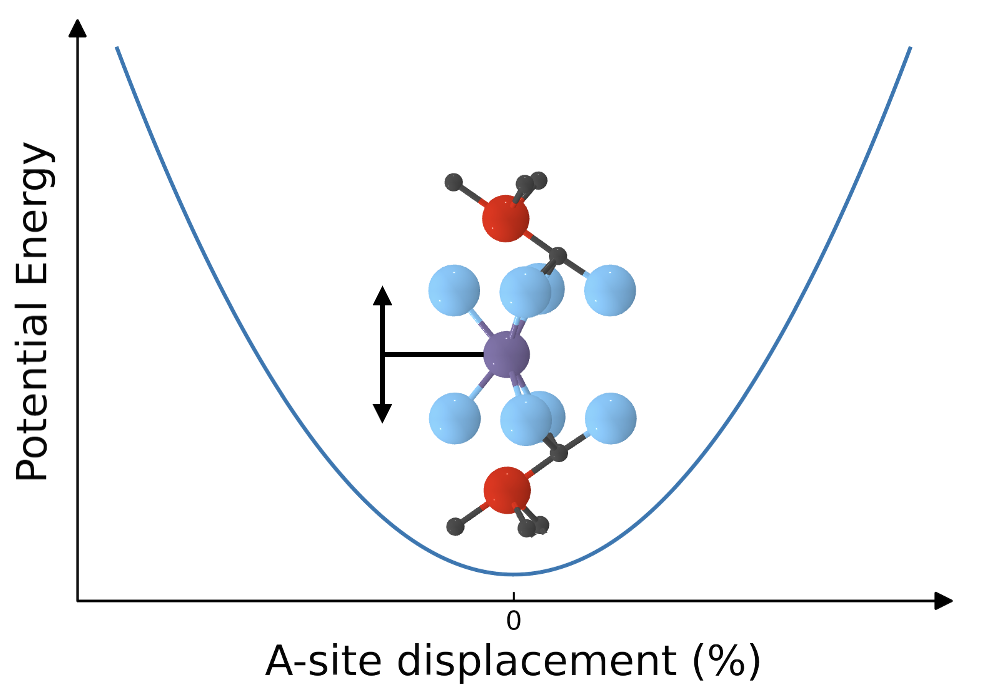}
    \caption{Schematic of an A-site sublattice perturbation in a representative double transition metal MAX phase unit cell and the corresponding potential energy change. The A-site atom is shown in purple, transition metal M-site atoms in light blue and red, and X-site atoms in black.}
    \label{fig:A_site_perturbation}
\end{figure}

The curvature of the potential well was estimated using a Taylor series expansion about the equilibrium geometry, with the unperturbed structure taken as the zeroth-order reference point. A schematic illustration of this procedure is shown in Fig.~\ref{fig:A_site_perturbation}. The finite difference expansion is given by:

\begin{align}
    E(\pm \delta) &= \sum_{n=0}^{\infty} \frac{E^{(n)}(0)}{n!} (\pm \delta)^n \label{eq:series} \\
    \implies E(\delta) &= E(0) + E'(0)\delta + \frac{E''(0)}{2}\delta^2 + \mathcal{O}(\delta^3) \label{eq:delta} \\
    \implies E(-\delta) &= E(0) - E'(0)\delta + \frac{E''(0)}{2}\delta^2 + \mathcal{O}(\delta^3) \label{eq:minusdelta}
\end{align}

Assuming that the energy above hull is minimized at $E(0)$, higher-order terms can be neglected ($\mathcal{O}(\delta^3) \approx 0$). Adding Eqs.~(\ref{eq:delta}) and (\ref{eq:minusdelta}) cancels the first derivative term, giving:
\begin{align}
    E(\delta) + E(-\delta) &= 2E(0) + E''(0)\delta^2 \\
    \implies E''(0) &= \frac{E(\delta) + E(-\delta) - 2E(0)}{\delta^2}
\end{align}

\begin{figure}[h!]
\centering
\includegraphics[width=\columnwidth]{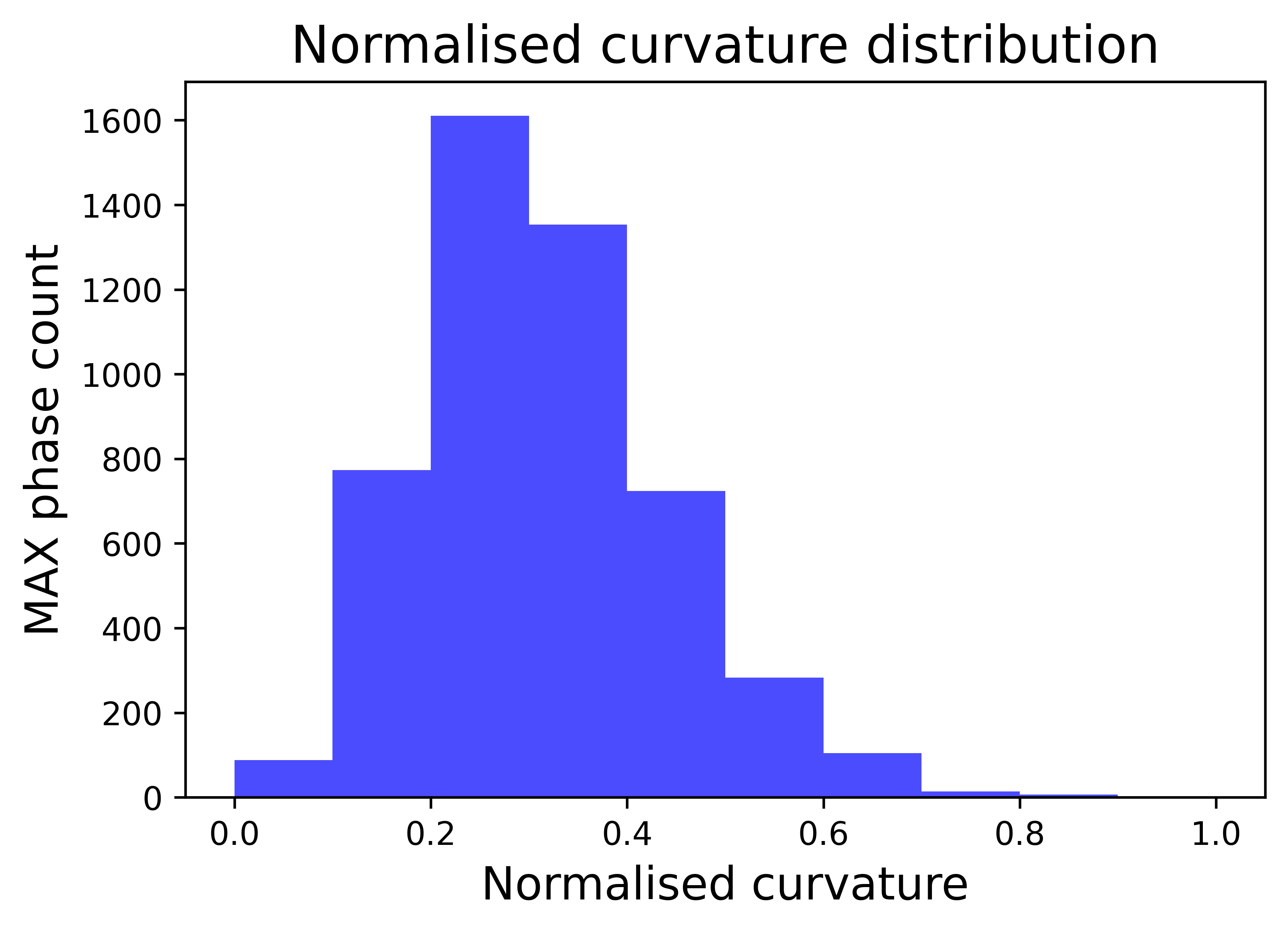}
\caption{Histogram displaying the min-max normalised A-site well curvature values against MAX phase count within the training dataset.}
\label{fig:normalised_curvature_distribution}
\end{figure}

This was then normalised across the dataset using min-max scaling to give a value between 0 and 1 as was the case in the previously considered potential MXene derivative count dimension. A histogram displaying this distribution can be seen in Fig.~\ref{fig:normalised_curvature_distribution}.

Further justification for this method was provided by the observation that, in all cases, the mapped potential energy surface was convex. Moreover, taking the noise level of MACE to be approximately 1–5 meV/atom,\cite{batatia2023} the signal exceeded the noise in every instance when considering the upper bound of 5 meV/atom. Additionally, a harmonic regime was preserved in over 80\% of cases, supporting the conclusion that this procedure reliably probed the bottom of the potential energy well.

\subsection{Conditional generation}

We assess the impact of conditional prompting on the properties of the generated structures. In this experiment we do not specify the composition of the prompt, but only use the conditioning vector to guide generation. Directly evaluating the effect of the first dimension, the MXene derivative count, was challenging, as this value is purely statistical and would require knowledge of the complete set of possible MXene derivatives at the point of generation to assess fidelity. By contrast, the second dimension, A-site well curvature, can be evaluated more directly.

\begin{figure}[h!]
    \centering
    \includegraphics[width=\columnwidth]{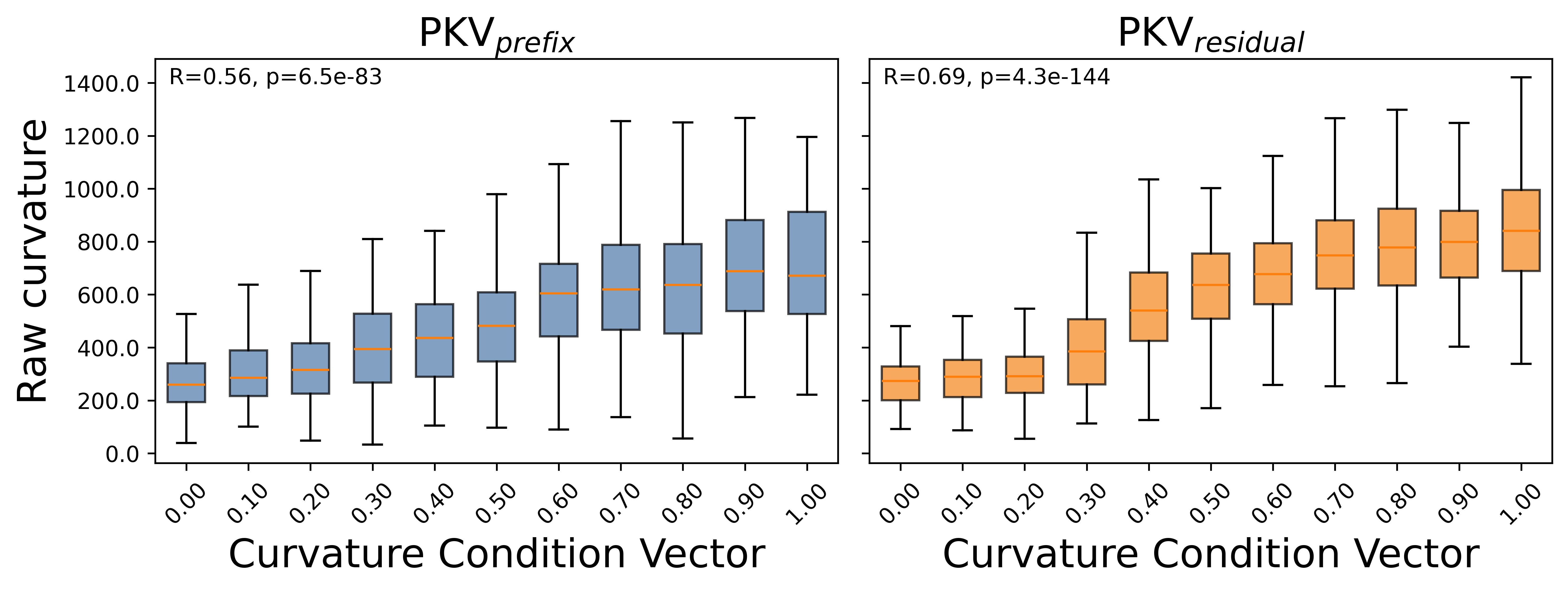}
    \caption{Plots showing the calculated curvature (arbitrary units) across a sweeping A-site well condition vector, with constant MXene derivative count dimension.}
    \label{boxplots}
\end{figure}

To this end, we consider the 1,000 structures generated under non-compositional prompting for each of the two conditioning mechanisms PKV$_{prefix}$ and PKV$_{residual}$ models, using a fixed MXene derivative count condition of 0.95 while sweeping the A-site well curvature from 0.00 to 1.00 in increments of 0.10. The resulting A-site well curvatures were calculated as described in the methods and plotted against the condition vector values. The results are shown in Fig.~\ref{boxplots}.

We observe a clear upward trend in A-site curvature with increasing values of the A-site condition vector. For PKV$_{prefix}$, the Pearson correlation is $r = 0.56$ ($p = 6.5 \times 10^{-83}$), while for PKV$_{residual}$ it is $r = 0.69$ ($p = 4.3 \times 10^{-144}$), detailing very high significance.

The disparity in $r$ values can be understood by considering the way missing values are handled by each model throughout the dataset. The imputation required to handle some structures in the PKV$_{prefix}$ model may be detracting from physically meaningful properties associated with some structures and thus diluting the meaning of the condition vector itself.

\begin{figure}[h!]
  \centering
  \includegraphics[width=\columnwidth]{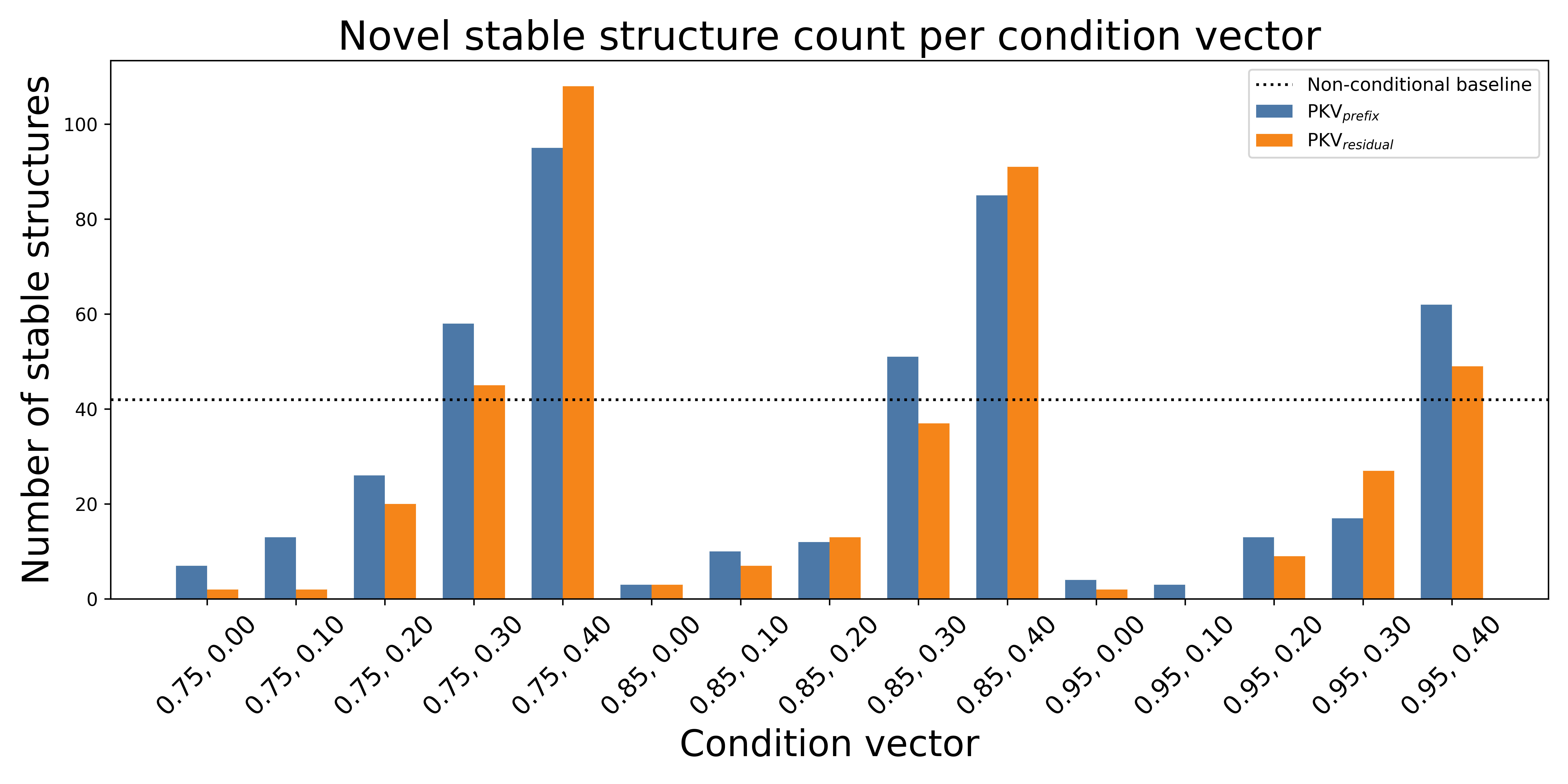}
  \caption{Bar chart detailing novel stable structure generation counts with respect to condition vector under non-compositional prompting. The dashed line indicates the non-conditional model baseline. Prompting performed within a condition vector space biased toward high potential MXene derivative counts and low A-site well curvature.}
  \label{fig:non_compositional}
\end{figure}

We now concentrate on generation in the region corresponding to high MXene derivative counts and low A-site well curvature. This region was defined as 0.75-0.95 (inclusive, step size 0.10) for the MXene derivative dimension and 0.0-0.40 (inclusive, step size 0.10) for the A-site well curvature dimension.
Within this defined sweep space, biased toward high MXene derivative counts and weak A-site binding, non-compositional prompting showed that the PKV$_{prefix}$ and PKV$_{residual}$ models generated 459 and 415 novel stable structures, respectively, compared to 644 from the baseline model. These results correspond to novel structure generation rates of 2.32\%, 2.10\%, and 3.25\% for PKV$_{prefix}$, PKV$_{residual}$, and the baseline, respectively.
It is evident, however, that relative to the non-conditional baseline, performance varies considerably across condition vectors. Some vectors outperform the baseline by a wide margin, while others under-perform. This variability is clearly illustrated in Fig.~\ref{fig:non_compositional}.
At (0.75, 0.40) and (0.85, 0.40), both models significantly outperformed the baseline, with odds ratios of 2.09-2.71 (p $<$ 0.001), indicating that conditioning under these settings more than doubled the likelihood of generating stable structures. By contrast, at (0.75, 0.30), neither PKV$_{prefix}$ nor PKV$_{residual}$ achieved statistical significance (95\% CI, p $<$ 0.05). At (0.95, 0.40), PKV$_{prefix}$ showed marginal significance (OR = 1.50, p = 0.028) while PKV$_{residual}$ did not. From these results we can conclude that generation rates are highly dependent on the specific condition vector employed.

Overall we can clearly see that conditioning not only has a meaningful impact on specific structural features but at specific condition vectors it is encouraging the model to explore a space capable of yielding more structurally novel and stable compounds than its non-conditioned counterpart. 
It can also be observed that novel stable structure generation rates increase with rising A-site well curvature, a trend that aligns with chemical intuition, i.e. a more tightly bonded A-site is more likely to yield a more stable structure. This highlights the delicate balance that needs to be achieved to propose stable, but easily exfoliated structures. The conditioning vector sweep was not extended beyond this range, as the purpose of the experiment was to focus on a region that was more likely to yield ``MXeneable" structures, not to generate arbitrary novel structures.

\subsection{Elemental prevalence}

Having established that conditioning can steer structural features and improve rates of novel structure generation, we next examine the elemental distribution across the previously defined set of novel stable structures. Because this set has already been filtered for metastability ($E_{hull} < 0.1$ eV/atom) and stoichiometric compliance (M$_{n+1}$AX$_{n}$), the analysis that follows concerns relative thermodynamic favourability and elemental generation frequency within the stable region.

\begin{figure*}[t]
    \centering
    \includegraphics[width=1.7\columnwidth]{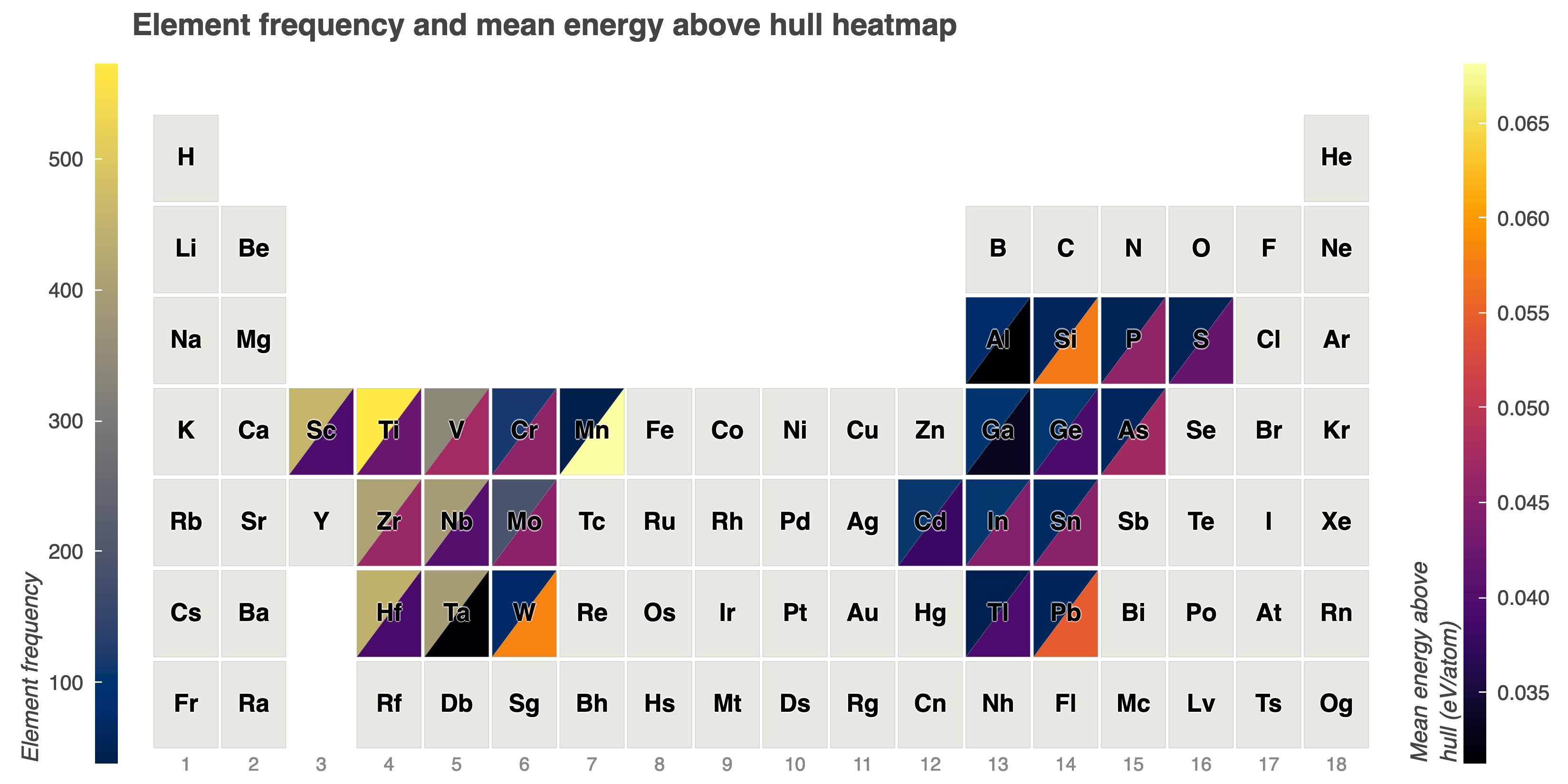}
    \caption{Element frequency heatmap for the set of novel stable structures showing element usage weighted by stoichiometry (upper left corner) and the mean energy above hull (lower right corner).}
    \label{fig:non_compositional_heatmap}
\end{figure*}

Figure~\ref{fig:non_compositional_heatmap} shows the frequency of each element in this set, together with the average $E_{hull}$ of all structures containing that element. To avoid skewing colour intensities, the X-site elements were excluded, but as expected, all resolved to either carbon or nitrogen. Element counts were extracted from the reduced formula to provide stoichiometrically weighted prevalence across generations. Additionally, a stoichiometry-weighted frequency restricted to structures with $E_{hull} < 0.05$ eV/atom is reported to distinguish elements that dominate the most thermodynamically favourable half of the (meta-) stable region.

Figure~\ref{fig:non_compositional_heatmap} shows that Ti is the most prevalent M-site element, appearing with a stoichiometry-weighted frequency of 573 across 106 distinct structures and a mean $E_{hull}$ of $0.042$ eV/atom. Ti also appears frequently in the region corresponding to $E_{hull} < 0.05$ eV/atom, contributing a weighted frequency of 238 structures - the highest of any element. The prominence of Ti is expected, as Ti-based MAX phases were among the first discovered, widely studied, and synthesised, and their derivatives - particularly \ce{Ti3C2T$_x$} - were also among the first MXenes reported.\cite{alam2024,gogotsi2019} This indicates that the model is successfully recovering experimentally observed trends. Elemental prevalence across group 4 and 5 is also relatively consistent, with Sc (454), Hf (447), Zr (408), Ta (396), Nb (395) and V (341) all exhibiting substantial representation, however, the remaining M-site elements fall off sharply, where the next highest prevalence is observed in Mo which only notes a count of 189. Notably, Ta exhibits the lowest mean $E_{hull}$ of any M-site element ($0.032$ eV/atom) despite ranking fifth by frequency, suggesting that Ta-containing MAX phases, while less frequently generated, are potentially more thermodynamically favourable when they do appear under this conditioning regime. Hf also shows a comparatively low mean $E_{hull}$ ($0.039$ eV/atom) and the second highest count of $E_{hull} < 0.05$ eV/atom (189). Hf-containing MAX phases are however, less common in the literature,\cite{lapauw2016} suggesting Hf-based compositions may be promising candidates for future study. Among the less prevalent M-site elements, W and Mn exhibit the highest mean $E_{hull}$ values ($0.058$ and $0.068$ eV/atom, respectively) and the lowest fractions of structures with $E_{hull} < 0.05$ eV/atom (30 and 6), indicating that while stable W and Mn containing MAX phases exist within the generated set, they are less favoured by the imposed conditioning regime.

For the A-site, reduced prevalence is observed, though this is expected given stoichiometric weighting (A-site elements contribute one count per formula unit in M$_{n+1}$AX$_n$). Cd is the most frequent A-site element (109), followed closely by In (105), Ge (103), and Ga (102). Al, despite a lower overall frequency of 74, exhibits the lowest mean $E_{hull}$ of any A-site element ($0.031$ eV/atom), followed by Ga ($0.033$ eV/atom). Both Al and Ga also show high representation for cases with $E_{hull} < 0.05$ eV/atom, noting stoichiometry-weighted frequencies of 38 and 54 respectively. The thermodynamic favourability of Al-containing MAX phases is consistent with experimental observations of Al as a common A-site element in both synthesised MAX phases and MXene-yielding precursors,\cite{dahlqvist2025} while Ga, although less studied, features regularly in the literature,\cite{dahlqvist2024} again indicating consistency between the model and well-established trends. Among other A-site elements, Ge achieves a mean $E_{hull}$ of $0.039$ eV/atom with the highest frequency of A-site elements possessing $E_{hull} < 0.05$ eV/atom (43), while Cd, despite its high overall prevalence, shows greater variance ($\sigma = 0.050$ eV/atom) and a median $E_{hull}$ of $0.052$ eV/atom, indicating a broader spread of output structures. At the less thermodynamically favourable end, Si and Pb exhibit mean $E_{hull}$ values of $0.057$ and $0.055$ eV/atom, with commensurate lower counts of $E_{hull} < 0.05$ eV/atom (12 and 14, respectively).

Since all structures in this set already satisfy the $0.1$ eV/atom metastability criterion, the relevant distinction lies not in whether elements pass a threshold but in the consistency of their stability. We are able to assess this by considering the standard deviation of $E_{hull}$ values for each element.  Elements such as Ta ($\sigma = 0.047$ eV/atom) and Sc ($\sigma = 0.047$ eV/atom) show relatively high variance, meaning that while their mean stability is favourable, individual compositions span a wider range. By contrast, Mn, despite having the highest mean $E_{hull}$ among M-site elements, exhibits the lowest standard deviation ($\sigma = 0.021$ eV/atom), indicating that, CrystaLLM$-\pi$ under the considered conditioning regime, can generate consistent structures within a tight thermodynamic range. These observations suggest that targeting high-frequency, low-variance elements - such as Ti, Hf, and Nb among M-sites, and Al and Ga among A-sites - may offer the most reliable generative route to thermodynamically favourable novel MAX phases.

Importantly, the fact that every M-site and A-site element in this set exhibits a mean $E_{hull}$ well below $0.1$ eV/atom confirms that viable candidates exist across the full range of conventional MAX phase chemistry, even if their relative abundance and thermodynamic favourability vary.

\subsection{Novel composition generation and DFT validation}

Across all non-compositional generations produced by the two conditional models under all condition vectors, a set of out-of-sample compositions were identified that appeared to correspond to stable phases. These compositions were absent from both the base training set and the fine-tuning set. A list of these compositions, together with predicted bulk moduli and shear moduli, is provided in Table~\ref{tab:max_properties}. 

\begin{table*}
\centering
\caption{Novel compositions predicted by CrystaLLM$-\pi$ with stability validated using MACE, across both conditional models and all condition vectors alongside bulk moduli and shear moduli predicted by ALIGNN.}
\label{tab:max_properties}
\begin{ruledtabular}
\begin{tabular}{lddd}
\textrm{Formula} & \multicolumn{1}{c}{\textrm{$E_{hull}$ (eV/atom)}} & \multicolumn{1}{c}{\textrm{Bulk moduli, B (GPa)}} & \multicolumn{1}{c}{\textrm{Shear moduli, G (GPa)}} \\
\colrule
TaMo$_2$GeC$_2$    & 0.00 & 236.47 & 93.85 \\
VW$_2$GeC$_2$      & 0.10 & 243.41 & 84.31 \\
ZrW$_2$SiC$_2$     & 0.10 & 246.42 & 86.74 \\
Zr$_3$GeC$_2$      & 0.03 & 167.07 & 72.30 \\
Hf$_2$ScInN$_2$    & 0.03 & 174.75 & 81.06 \\
HfZr$_2$CdC$_2$    & 0.00 & 158.81 & 60.73 \\
Sc$_2$WPbC$_2$     & 0.09 & 101.79 & 70.25 \\
%Hf$_2$TiGeN$_2$    & 0.09 & 181.12 & 79.91 \\
Sc$_2$TaInN$_2$    & 0.10 & 136.03 & 66.65 \\
ZrTi$_2$SnC$_2$    & 0.04 & 169.27 & 75.18 \\
%Ti$_2$NbInN$_2$    & 0.09 & 176.57 & 79.40 \\
ScTi$_2$CdC$_2$    & 0.09 & 153.53 & 64.02 \\
\end{tabular}
\end{ruledtabular}
\end{table*}

The ratio of shear to bulk moduli (G/B) generally falls between 0.4 and 0.6, indicative of ductility commonly associated with MAX phases.\cite{das2023} Additionally, structures listed in this table have, in some capacity, been included within the search space for MAX phases in prior literature. However, only one phase has been explicitly investigated to date: Zr$_3$GeC$_2$, which has been verified as theoretically stable through DFT calculations performed at the PBE level using the generalized gradient approximation (GGA).\cite{das2023} To the best of our knowledge, the remaining structures have not been explicitly investigated.

\begin{table*}
\centering
\caption{DFT relaxation calculations performed at the PBE level for the established novel stable compositions as predicted by CrystaLLM$-\pi$ and MACE.}
\label{tab:max_candidates}
\begin{ruledtabular}
\begin{tabular}{ldddd}
\textrm{Formula} &
\multicolumn{1}{c}{\textrm{$E_{hull}$ (eV/atom)}} &
\multicolumn{1}{c}{\textrm{Formation energy (eV/atom)}} &
\multicolumn{1}{c}{\textrm{Energy (eV)}} &
\multicolumn{1}{c}{\textrm{Band gap (eV)}} \\
\colrule
TaMo$_2$GeC$_2$    &  0.020 & -0.256 & -116.33 & 0.00 \\
VW$_2$GeC$_2$      &  0.233 &  0.011 & -116.03 & 0.00 \\
ZrW$_2$SiC$_2$     &  0.112 & -0.283 & -120.23 & 0.00 \\
Zr$_3$GeC$_2$      & -0.037 & -0.842 & -107.55 & 0.00 \\
Hf$_2$ScInN$_2$    &  0.047 & -1.391 & -106.59 & 0.00 \\
HfZr$_2$CdC$_2$    & -0.034 & -0.664 & -100.83 & 0.07 \\
Sc$_2$WPbC$_2$     & -0.011 & -0.428 & -100.72 & 0.00 \\
%Hf$_2$TiGeN$_2$    &  0.136 & -1.358 & -113.07 & 0.00 \\
Sc$_2$TaInN$_2$    &  0.162 & -1.240 & -101.33 & 0.00 \\
ZrTi$_2$SnC$_2$    &  0.062 & -0.724 & -102.30 & 0.00 \\
%Ti$_2$NbInN$_2$    &  0.176 & -1.091 & -102.29 & 0.00 \\
ScTi$_2$CdC$_2$    &  0.170 & -0.466 &  -88.60 & 0.00 \\
\end{tabular}
\end{ruledtabular}
\end{table*}

If we tighten our metrics further and define thermodynamic stability as structures lying less than 0.050 eV/atom above the convex hull, we find that 5 out of the 10 candidates satisfy our criterion as seen in Table~\ref{tab:max_candidates}. This indicates that the model and energy evaluation metrics achieve $\sim 50$\% fidelity with respect to the specified level of DFT theory when considering out-of-sample composition generation. Band gap calculations confirm that all candidates are metallic in character, another property commonly associated with MAX phases. In the case of the HfZr$_2$CdC$_2$ we might assume the small deviation from 0 eV is simply an artefact of calculation. 

Considering the origin of these novel compositions, some emerge from regions of the condition vector space associated with a greater number of potential MXene derivatives and reduced A-site well curvature. As defined by the PKV$_{residual}$ and/or PKV$_{prefix}$ models, Hf$_2$ScInN$_2$ corresponds to a condition vector of (0.95, 0.40), Sc$_2$PbWC$_2$ to (0.95, 0.30), HfZr$_2$CdC$_2$ to (0.75, 0.30), TaMo$_2$GeC$_2$ to (1.00, 0.50), and Zr$_3$GeC$_2$ to (0.50, 0.75). Notably, TaMo$_2$GeC$_2$ and Zr$_3$GeC$_2$ were generated outside the primary region of condition vector space considered in this work. These arose from preliminary exploratory screening conducted early in the experimental workflow. Despite being outside our target conditioning space, these compositions were retained due to their compositional novelty relative to the training set. 

Beyond our energetic filtering criteria, we also identified ZrTi$_2$SnC$_2$ from the condition vector (0.85, 0.30). Despite this composition failing our more stringent screening, the discrepancy is small and the compound arises from a desirable region of conditioning space and thus may represent a promising candidate for future investigation if structural relaxation techniques could be applied to satisfy stricter stability requirements.

While it reflects positively on model capability that feasible out-of-sample compositions can be generated within the targeted vector space, these calculations alone are insufficient to confirm that the resulting structures are able to be etched into MXenes. Considering the most common top-down synthesis route via HF acid etching, Björk \textit{et al.}\cite{bjork2023} demonstrated that the chemical exfoliation process can be effectively understood by evaluating the free energy difference associated with removing A- and M-site elements, under the assumption that these species do not dissolve as a consequence of water splitting. This approach could therefore serve as a valuable subsequent screening step in future work to assess the etchability and exfoliation potential of generated MAX phases.

\subsection{Low shot generation - Borides}

A total of 1,670 MAB phase generations produced 619 valid boride CIFs, of which 28 were classified as novel and stable under our defined metrics. These corresponded to \ce{Ta4Si4B4}, \ce{Nb4Ga4B4}, \ce{Nb4Si4B4}, \ce{V4Si4B4}, \ce{Cr4Si4B4}, and \ce{Mo4Ga4B4}. While these compositions have been previously reported, none have been documented in the crystal structures generated here to the best of our knowledge, thus qualifying as novel by our metrics.

DFT calculations on these structures revealed that 23 of the 28 candidates were metastable ($E_{\mathrm{hull}} \lesssim 0.051$~eV/atom) under the defined level of DFT theory, and all exhibited a band gap metallic in character. While this demonstrates the model’s capacity to recover physically reasonable structures in a low-shot regime, the diversity of the generated outputs remains limited: only four unique compositions out of the original six (\ce{Mo4Ga4B4}, \ce{Nb4Si4B4}, \ce{Ta4Si4B4}, and \ce{V4Si4B4}) satisfied the DFT-screening and all adopt a 1:1:1 MAB stoichiometry with the same \textit{Cmcm} space group. In this regime, conditional and non-conditional generation rates closely track one another across the explored chemical space, with neither approach exhibiting a clear advantage. This convergence suggests that, under conditions of limited data availability, the model is unable to learn sufficiently strong or distinct structure--property relationships for conditional signals to meaningfully influence generation. Such behaviour likely reflects a combination of the aggressive learning rate employed during fine-tuning and the sparse representation of MAB-type structures in the training set. To more effectively integrate MAB phase generation into the broader workflow, future efforts should therefore focus on expanding and balancing the dataset to encompass a wider range of canonical MAB phases and stoichiometries, thereby alleviating the implicit constraints imposed by the low-shot regime.

\section{Conclusion}

We have introduced a generative model, tailored to design new MAX materials. The model, based on CrystaLLM$-\pi$ employs conditional autoregressive generation to propose new crystal structures biased towards having MAX structure and being amenable to exfoliation to MXene derivatives. To achieve this conditioning we developed two heuristics of MXene forming likelihood, one statistically derived from literature reports and one physically motivated based on bond strength. We demonstrate that these conditions can drive the underlying generative process to propose new materials that are physically more likely to have easy routes to exfoliation. We then further apply the model to propose new MAX materials and also specifically target the less explored MAB chemical space. We select some of the most stable novel compositions (as evaluated using the MACE machine learned interatomic potential) and test their stability and properties using DFT, identifying five new (meta-)stable phases that to our knowledge have not been explored as promising MAX/MXene materials previously. We also use the model to expand the space of potential MAB materials - proposing 23 new structurally unique (meta-)stable materials from DFT calculations. 

\section{Methods}

\subsection{General workflow}

\begin{figure}
    \includegraphics[width=\columnwidth]{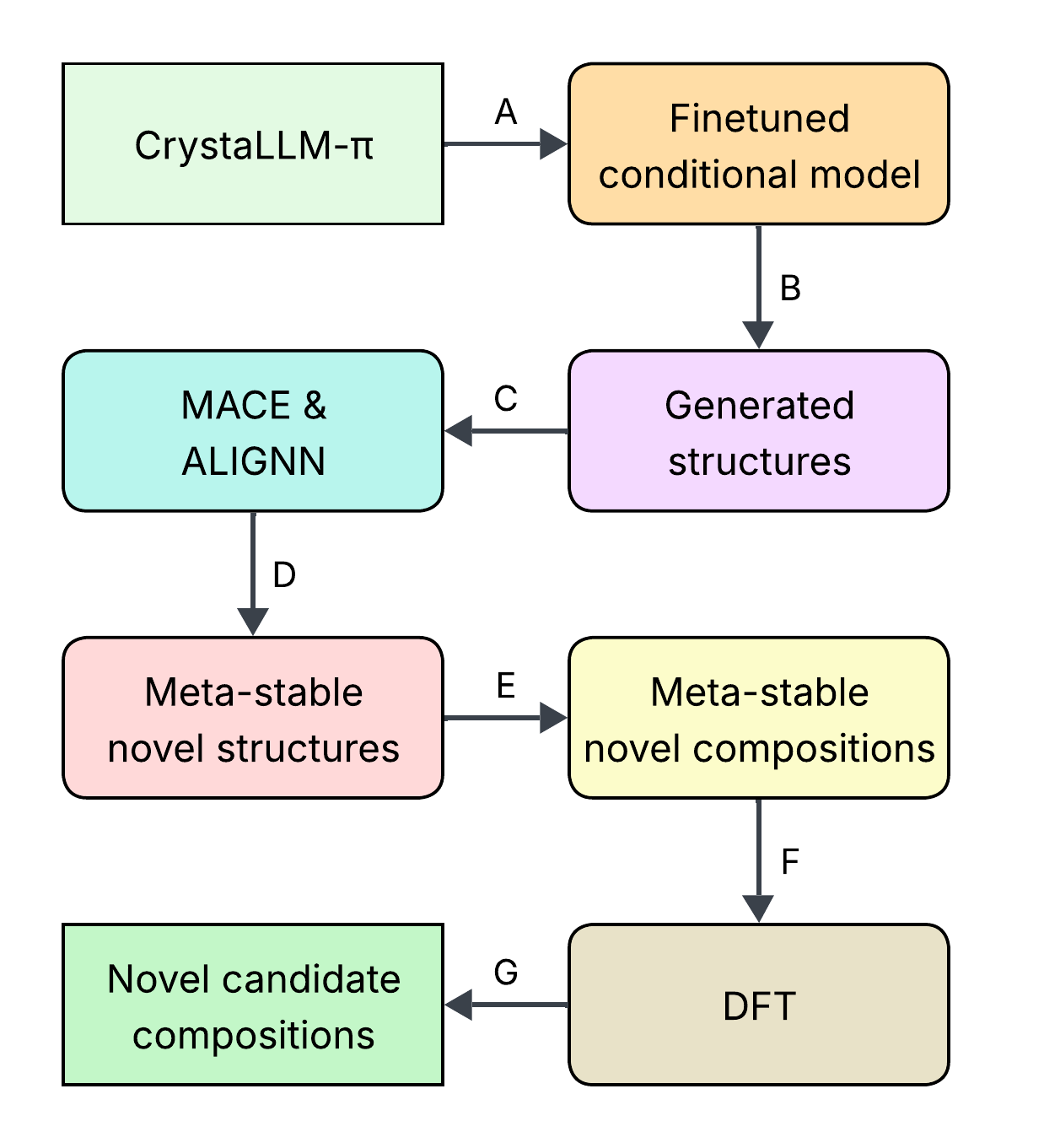}
    \caption{Full workflow for the conditional generative process.}
    \label{fig:workflow_crystallmpi}
\end{figure}

The complete investigative workflow is summarised in Fig.~\ref{fig:workflow_crystallmpi}. 
In step A, we begin with a base model pre-trained on a diverse general corpus of approximately $4.25$ million crystal structures,\cite{lemat_bulkunique_2024} which is subsequently fine-tuned on a MAX-phase-specific dataset.\cite{nykiel2023}

In step B, this refined model is used for structural generation via a systematic sweep of the predefined condition vector, thereby exploring the target region of materials design space. At this stage, generation may proceed under either compositional or non-compositional prompting. 

Step C comprises a multi-stage filtering process in which all generated structures are first validated for compliance with CIF semantic rules, internal structural consistency, and basic physical plausibility. Structures that pass this screening are subsequently evaluated using the MACE and ALIGNN models to assess thermodynamic stability and predict relevant materials properties, enabling efficient elimination of unphysical or thermodynamically unstable candidates.

In step D, candidates that are both physically plausible and thermodynamically reasonable are screened for structural novelty against both the base dataset and the MAX-phase fine-tuning set, ensuring that only stable and structurally unique outputs are retained. Structural novelty is assessed through symmetry-based screening, in which candidates are considered distinct only if they differ beyond defined tolerances: a 0.1~\AA{} criterion for symmetry-equivalent positions, a 0.3~\AA{} site displacement tolerance, a 20\% variation in lattice parameters, and a $5^\circ$ angular tolerance. 

These meta-stable, structurally novel candidates are subsequently filtered based on chemical composition, with novelty again assessed relative to the full training and fine-tuning corpus (step E).

In step F, the remaining meta-stable and compositionally novel structures are subjected to density functional theory (DFT) calculations, providing a more stringent evaluation of thermodynamic stability. 
Finally, in step G, the candidates are filtered according to their DFT-derived energies and cross-referenced with their originating regions of condition vector space, yielding a final set of compositionally novel structures that are likely to exhibit desirable MAX-phase characteristics.

\subsection{Data Collection}

This study makes use of a dataset of 8,591 double transition-metal MAX phase structures generated using DFT at the PBE level.\cite{nykiel2023} These structures underwent ionic relaxations with a stopping criterion of $10^{-5}$ eV and cell relaxations with a stopping criterion of $10^{-6}$ eV, with convergence ensured by employing a plane-wave cutoff energy of 550 eV and a k-point grid of $18 \times 18 \times 3$. Within the domain of double transition-metal MAX phases, this database has constituted the largest collection of its kind since 2023.

\begin{figure}[h!]
    \centering
    \includegraphics[width=\columnwidth]{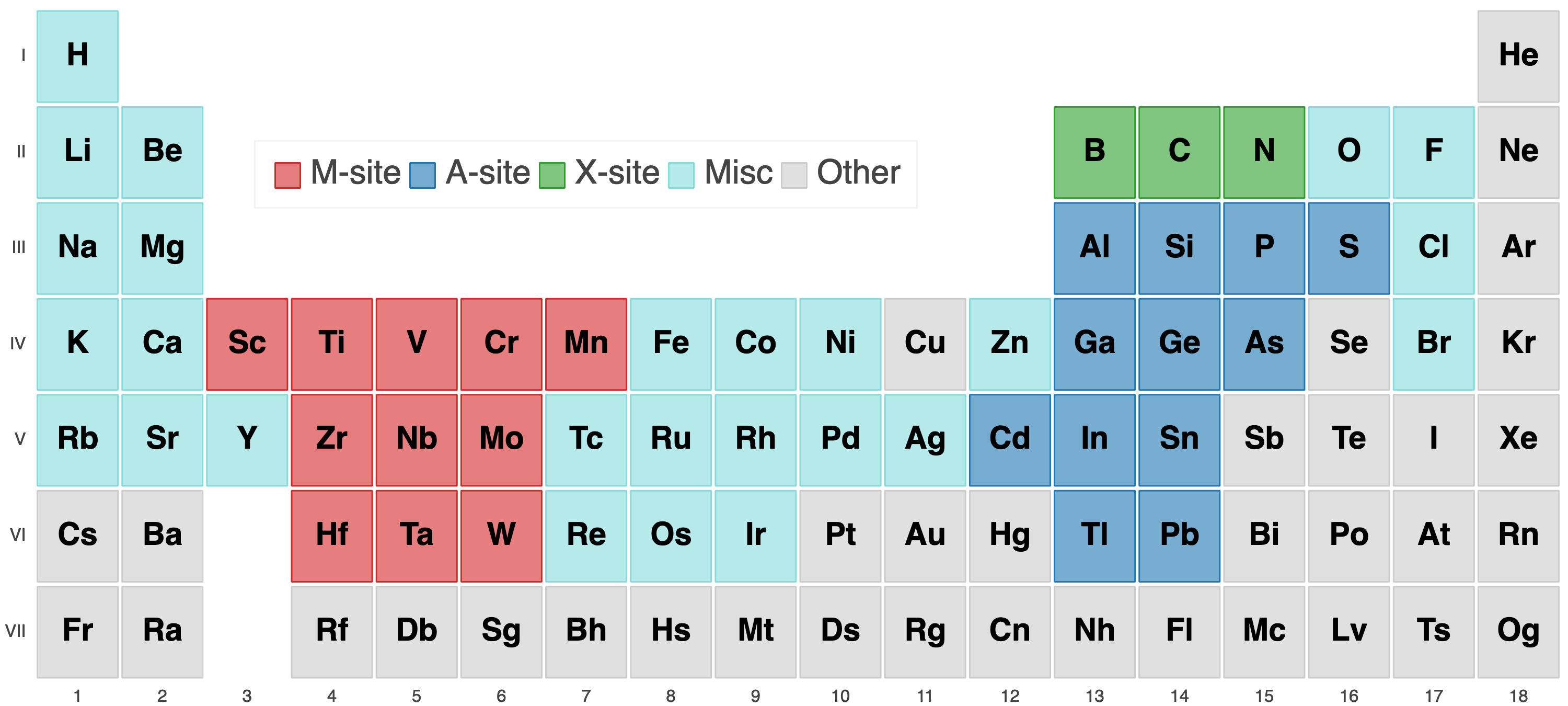}
    \caption{The elemental space of the modified fine-tuning dataset including miscellaneous elements that appear in the MAX/MAB-like set.}
    \label{fig:finetuning_elemental_space}
\end{figure}

To complement this, an additional set of 166 compositionally unique MAB and MAX-like structures was obtained from the Materials Project and NOMAD databases. These were filtered based on theoretical stability, exclusion of rare-earth elements, and assignment to one of the following space groups: P6$_3$/mmc, P6$_3$mc, P6$_3$/mcm, R$\bar{3}$m, Cmcm, Pnma, Cmmm, P$\bar{6}$m2, P$\bar{3}$m1, and P6$_3$. This step enriched the dataset with additional chemical and structural diversity, particularly for boride compounds (104 in total, all classified as stable by the Materials Project, NOMAD, and the stability metrics employed in this study) as well as MAX-like compounds. This was necessary because the original double transition-metal MAX phase dataset contained only carbides and nitrides, and therefore did not capture or even vaguely reference the distinct chemistry of MAB phases (\ce{M2AB2}, \ce{M3AB4}, \ce{M4AB6}, \ce{M4AB4}, \ce{M2A2B2}).\cite{dahlqvist2024} Overall, 6,179 species in this dataset emerged as structurally unique under our metrics and were therefore passed forward to be used in model fine-tuning. The elemental space of the fine-tuning dataset is illustrated in Fig.~\ref{fig:finetuning_elemental_space}.

An additional dataset comprising 23,857 MXene phases, also generated using DFT at the PBE level (the aNANt functional materials database\cite{anant}), was also considered. This dataset was specifically used to define the MXene derivative count dimension of the condition vector.

\subsection{Thermodynamic Stability Evaluation}

Total energies were calculated using the MACE-MP (Message Passing Atomic Cluster Expansion) calculator,\cite{batatia2023} employing its default model pre-trained on Materials Project structure relaxation trajectories. To assess phase stability, formation energies of known materials from the Materials Project were used to construct a convex hull.

The convex hull provides a straightforward means of determining thermodynamic stability within a chemical system. It is constructed by plotting the formation energies of all known or theoretically predicted compounds against their compositions and connecting the lowest-energy points to form a ``hull," which represents the set of compositions stable against decomposition into other phases. Compounds that lie on the hull are thermodynamically stable, whereas those positioned above it are unstable and tend to decompose into combinations of phases residing on the hull.\cite{anelli2018}  

From this framework, the $E_{hull}$ metric was defined as the energy above the convex hull for each structure, calculated by comparing its formation energy against the most thermodynamically stable competing phases with the same composition and stoichiometry. All competing phase energies were computed using MACE-MP. Structures with $E_{hull}<0.1$ eV/atom were classified as meta-stable.\cite{jain2013}

\subsection{Property Key Value (PKV) conditioning}

Two conditioning methods were considered in this study. The first, denoted PKV$_{prefix}$, incorporates a learnable embedding of the condition vector directly into the attention mechanism, representing a stronger form of conditioning. The second, PKV$_{residual}$, provides a softer alternative in which two attention scores, one attending to the sequence and one to the conditioning information, are computed in parallel, where the next token prediction is obtained as a weighted sum of these contributions. This formulation allows PKV$_{residual}$ to naturally handle missing values in the condition vector.\cite{bone2025discoveryrecoverycrystallinematerials}

Missing values were treated using two distinct strategies. The first model, PKV$_{residual}$, is inherently capable of handling missing entries in the condition vector by dynamically adjusting how information is injected into the model's attention mechanism. The second model, Property Key Value (PKV$_{prefix}$), required explicit imputation, which was carried out by substituting the normalised mean value for each missing instance.

Owing to the structural and compositional differences between MAB/MAB-like and canonical MAX phases, all condition vectors were imputed for boride structures during training. Although sub-optimal for low-shot generation, this approach enables uniform algorithmic treatment of all training structures, ensuring consistency across this experimental framework. Independent MAB phase conditioning may therefore, be a valuable direction for future investigation.

\subsection{Fine-tuning and hyperparameter optimisation}

The fine-tuning dataset was divided into training, validation, and test sets in an 80:10:10 ratio. Borides were withheld from the test set to maximize their representation in the training and validation sets under a low-shot regime. The test set was used solely to verify outputs prior to exploratory MAX phase generations and is not considered further in this work. Hyperparameter tuning was conducted over a restricted search space by sweeping a grid of 20 plausible configurations, varying the learning rate, prefix token, and hidden condition size. The optimal model was selected on the basis of the lowest validation loss achieved.

Three models were subsequently fine-tuned: the PKV$_{prefix}$ model, the PKV$_{residual}$ model, and a baseline model without a condition vector. Where applicable, all models were trained under identical settings to ensure a fair and consistent comparison.

\subsection{Structure generation}

Structure generations were carried out using both compositional (MAB phases) and non-compositional prompting. For the conditioning models, a series of condition vectors was additionally swept across both prompting modes.

\subsubsection{Non-compositional prompting}

Multiple investigations were carried out under this regime. First, 1,000 structures were generated for each model type at a fixed MXene derivative count condition of 0.95, while sweeping the A-site well curvature condition from 0.00 to 1.00 in increments of 0.10. Raw well curvature values were then calculated for each generated structure to assess the influence of the A-site well curvature condition on generation outcomes. In this instance, structures were filtered only for validity meaning that energy ranges and novelty were not considered. Significance was again assessed at the 5\% level using a two-sided Pearson correlation test, with the null hypothesis that there is no correlation between the A-site well curvature condition and the measured curvature of the generated structures.

Non-compositional prompts were swept across a narrower range: 0.75 to 0.95 (inclusive, step size 0.10) for the MXene derivative dimension, and 0.0 to 0.4 (inclusive, step size 0.10) for the A-site well curvature dimension. This produced 15 distinct condition vectors. These ranges were chosen pragmatically, as they correspond to regions where many potential MXene derivative candidates are defined and where weak A-site binding is suggested by low curvature values. Structures from this sweep were then filtered on the basis of novelty, validity, and metastability ($E_{hull} < 0.1$ eV/atom). See Table~\ref{tab:cv_interpretation} for a qualitative interpretation of what these condition vectors mean.

\begin{table}[h!]
\centering
\caption{Examples of conditioning vectors and their qualitative interpretation.}
\label{tab:cv_interpretation}
\begin{ruledtabular}
\begin{tabular}{c l}
\textbf{Condition Vector} &
\textbf{Qualitative Interpretation} \\
\colrule
(0.75, 0.00) &
\parbox[t]{0.55\columnwidth}{%
Lower MXene-derivative score within the swept region and
weakest A-site binding (lowest well curvature). } \\
(0.85, 0.20) &
\parbox[t]{0.55\columnwidth}{%
Intermediate MXene-derivative score with moderate A-site
well curvature in the explored range. } \\
(0.95, 0.40) &
\parbox[t]{0.55\columnwidth}{%
Highest MXene-derivative score within the swept region and
strongest A-site binding (highest well curvature). } \\
(0.95, 0.00) &
\parbox[t]{0.55\columnwidth}{%
Highest MXene-derivative score within the swept region combined with
weakest A-site binding (lowest well curvature). } \\
\end{tabular}
\end{ruledtabular}
\end{table}

In this setup, 19,800 structures were generated for each conditional model type, yielding 1,320 structures per condition vector. For comparison, the baseline model generated 19,800 structures without conditioning. 

To assess the significance of novel stable structure generation rates, in both cases a Fisher's exact test was performed at the 5\% significance level, comparing the performance of individual condition vectors for both model types against the baseline model. The null hypothesis was that the use of a specific condition vector does not alter the rate of novel stable structure generation relative to the baseline model.

\subsubsection{MAB phases - Low shot generations}

Due to reduced prevalence of MAB phases in the fine-tuning set, boride structures accounted for only 104 of 6,179 entries (1.68\%), providing a suitable case study for assessing the low-shot generation capabilities of CrystaLLM$-{\pi}$ in this context. Faithful MAB phases represent an under-explored subclass of MAX-type structures, noted for adopting distinct structural motifs and stoichiometries.\cite{dahlqvist2024} These phases can be etched into MBenes analogously to MXenes and are emerging as promising alternatives for applications in batteries, supercapacitors, and catalysis.\cite{alameda2018}

To systematically investigate boron-containing compositional space within this low-data regime, a series of 167 unique compositional prompts was generated using a predefined set of elements, following the stoichiometric rules of MAB phases. Prompt compositions were validated using SMACT\cite{smact} to ensure charge neutrality and electronegativity balance. For each of the 167 prompts, 10 structures were generated, yielding a total of 1,670 generations. The generation process followed a prompting methodology where condition vectors ranging from 0.00 to 1.00 were swept in increments of 0.25 for each dimension, resulting in 25 distinct condition vector sets. This conditioning strategy was applied uniformly across both PKV$_{\text{prefix}}$ and PKV$_{\text{residual}}$ model variants, enabling direct comparison of their respective performance in the boride chemical space.

\subsection{Structure matching and Novelty Evaluation}

Structural novelty was assessed using the \texttt{pymatgen} StructureMatcher tool.\cite{pymatgen} Matching was performed with the following thresholds: 0.1 \AA\ for symmetry equivalence, 0.3 \AA\ for site equivalence, 20\% variation in lattice parameters, and a 5$^\circ$ variation in lattice angles. Structures exceeding all thresholds were designated as structurally novel.

Compositional novelty was determined by direct comparison of \texttt{pymatgen} reduced formula composition objects from the generated set against those present in the full training dataset.

\subsection{DFT calculations}

DFT calculations were performed using VASP with a plane wave cutoff of 520 eV. Structural relaxation used the PBE functional with the projector augmented wave (PAW) method. Electronic convergence was set to $10^{-6}$ eV and ionic convergence to 0.05 eV/\AA. Structural relaxations were carried out in three regimes: (i) relaxation of atomic positions (ISIF=2), (ii) relaxation of both atomic positions and lattice vectors (ISIF=3), and (iii) short refinement runs. The conjugate-gradient algorithm was used with Gaussian smearing ($\sigma = 0.05$ eV, ISMEAR$ = 0$). Final static total-energy calculations were performed using the tetrahedron method with Bl\"ochl corrections (ISMEAR = -5) and no relaxation of ions (IBRION=-1). \textit{k}-point meshes were automatically generated using Pymatgen's MPRelaxSet and MPStaticSet to ensure compatibility with the Materials Project database. The maximum relaxation cycles were capped at 20. 

\subsection{Acknowledgements}
KTB acknowledges funding from EPSRC (EP/Y014405/1, EP/Y000552/1, EP/Y028775/1 and EP/Y028759/1). JS was funded under the AWE - UCL Chief Scientist Short Project Scheme. We acknowledge the use of UCL Myriad High Performance Computing Facility (Myriad@UCL), and associated support services, in the completion of this work. We acknowledge Young HPC access which is partially funded by EPSRC (EP/T022213/1, EP/W032260/1 and EP/P020194/1). CB is funded by a UCL start-up package. 

\subsection{Author Contributions}
KTB and JS conceptualised the project and wrote the initial manuscript. CB provided technical supervision and reviewed the manuscript. EG and MD provided guidance on chemical systems and contributed to conceptualisation and development of the project and manuscript.

\subsection{Data Availability}
All data and code needed to recreate the results presented in this paper are available from \url{https://github.com/j-swaine/Conditional-Generation-of-MAX-Phases} (code) and \url{https://huggingface.co/datasets/Jamie1701/conditional-generative-models-max-phase} (data).

\bibliography{main}

@article{nykiel2023,
  author    = {Nykiel, K. and Strachan, A.},
  title     = {High-throughput density functional theory screening of double transition metal MXene precursors},
  journal   = {Scientific Data},
  year      = {2023},
  volume    = {10},
  pages     = {827},
  doi       = {10.1038/s41597-023-02755-2},
}

@article{walker2026carbon,
  title={The carbon cost of materials discovery: Can machine learning really accelerate the discovery of new photovoltaics?},
  author={Walker, Matthew and Butler, Keith T},
  journal={Materials Horizons},
  year={2026},
  publisher={Royal Society of Chemistry}
}

@article{dahlqvist2024,
title = {MAX phases – Past, present, and future},
journal = {Materials Today},
volume = {72},
pages = {1-24},
year = {2024},
issn = {1369-7021},
doi = {https://doi.org/10.1016/j.mattod.2023.11.010},
url = {https://www.sciencedirect.com/science/article/pii/S1369702123003577},
author = {Martin Dahlqvist and Michel W. Barsoum and Johanna Rosen}
}

@misc{smact,
  author = {Anthony Onwuli and Daniel W. Davies and Keith T. Butler and Adam J. Jackson and Jonathan M. Skelton and Kazuki Morita and Aron Walsh},
  note = {\ \unskip SMACT, Materials Design Group (WMD-group), Imperial College London, London, UK, 2016. \url{https://github.com/WMD-group/SMACT}}
}

@misc{pymatgen,
  author = {Shyue Ping Ong and William Davidson Richards and Anubhav Jain and Geoffroy Hautier and Michael Kocher and Shreyas Cholia and Dan Gunter and Vincent Chevrier and Kristin A. Persson and Gerbrand Ceder},
  note = {\\unskip pymatgen, Computational Materials Science, 2013, 68, 314--319. doi:10.1016/j.commatsci.2012.10.028. \url{https://github.com/materialsproject/pymatgen}}
}

@misc{anant,
  note = {\unskip aNANt,\url{https://anant.mrc.iisc.ac.in/apps/mxene},accessed September 2025}
}

@article{das2023,
  author  = {P. Das and N. Jahan and M. A. Ali},
  title   = {DFT insights into Nb-based 211 MAX phase carbides: Nb2AC (A = Ga, Ge, Tl, Zn, P, In, and Cd)},
  journal = {RSC Advances},
  year    = {2023},
  volume  = {13},
  pages   = {5538--5556},
  doi     = {10.1039/d2ra07468k}
}

@article{benchakar2020,
title = {One MAX phase, different MXenes: A guideline to understand the crucial role of etching conditions on Ti3C2Tx surface chemistry},
journal = {Applied Surface Science},
volume = {530},
pages = {147209},
year = {2020},
issn = {0169-4332},
doi = {https://doi.org/10.1016/j.apsusc.2020.147209},
url = {https://www.sciencedirect.com/science/article/pii/S0169433220319668},
author = {Mohamed Benchakar and Lola Loupias and Cyril Garnero and Thomas Bilyk and Cláudia Morais and Christine Canaff and Nadia Guignard and Sophie Morisset and Hanna Pazniak and Simon Hurand and Patrick Chartier and Jérôme Pacaud and Vincent Mauchamp and Michel W. Barsoum and Aurélien Habrioux and Stéphane Célérier}
}

@article{bao2021,
title = {Role of MXene surface terminations in electrochemical energy storage: A review},
journal = {Chinese Chemical Letters},
volume = {32},
number = {9},
pages = {2648-2658},
year = {2021},
issn = {1001-8417},
doi = {https://doi.org/10.1016/j.cclet.2021.02.012},
url = {https://www.sciencedirect.com/science/article/pii/S1001841721000668},
author = {Zhuoheng Bao and Chengjie Lu and Xin Cao and Peigen Zhang and Li Yang and Heng Zhang and Dawei Sha and Wei He and Wei Zhang and Long Pan and Zhengming Sun}
}

@article{jain2013,
  author    = {Anubhav Jain and Shyue Ping Ong and Geoffroy Hautier and Wei Chen and William Davidson Richards and Stephen Dacek and Shreyas Cholia and Dan Gunter and David Skinner and Gerbrand Ceder and Kristin A. Persson},
  title     = {Commentary: The Materials Project: A materials genome approach to accelerating materials innovation},
  journal   = {APL Materials},
  volume    = {1},
  number    = {1},
  pages     = {011002},
  year      = {2013},
  month     = {July},
  doi       = {10.1063/1.4812323}
}

@article{anelli2018,
  title = {Generalized convex hull construction for materials discovery},
  author = {Anelli, Andrea and Engel, Edgar A. and Pickard, Chris J. and Ceriotti, Michele},
  journal = {Phys. Rev. Mater.},
  volume = {2},
  issue = {10},
  pages = {103804},
  numpages = {8},
  year = {2018},
  month = {Oct},
  publisher = {American Physical Society},
  doi = {10.1103/PhysRevMaterials.2.103804},
  url = {https://link.aps.org/doi/10.1103/PhysRevMaterials.2.103804}
}

@article{alam2024,
  author    = {Alam, M. S. and Chowdhury, M. A. and Khandaker, T. and Hossain, M. S. and Islam, M. S. and Islam, M. M. and Hasan, M. K.},
  title     = {Advancements in MAX phase materials: structure, properties, and novel applications},
  journal   = {RSC Advances},
  year      = {2024},
  volume    = {14},
  number    = {37},
  pages     = {26995--27041},
  doi       = {10.1039/d4ra03714f},
  pmid      = {39193282},
  pmcid     = {PMC11348849},
  publisher = {Royal Society of Chemistry},
  url       = {https://doi.org/10.1039/d4ra03714f},
  note      = {Published Aug 27, 2024}
}

@article{gogotsi2019,
author = {Gogotsi, Yury and Anasori, Babak},
title = {The Rise of MXenes},
journal = {ACS Nano},
volume = {13},
number = {8},
pages = {8491-8494},
year = {2019},
doi = {10.1021/acsnano.9b06394},
    note ={PMID: 31454866},

URL = { 
    
        https://doi.org/10.1021/acsnano.9b06394
    
    

},
eprint = { 
    
        https://doi.org/10.1021/acsnano.9b06394
    
    

}

}

@article{lapauw2016,
  author = {Lapauw, Thomas and Tunca, Bensu and Cabioc'h, Thierry and Lu, Jun and Persson, Per O. {\AA}. and Lambrinou, Konstantina and Vleugels, Jozef},
  title  = {Synthesis of {MAX} Phases in the Hf--Al--C System},
  journal= {Inorganic Chemistry},
  volume = {55},
  number = {21},
  pages  = {10922-10927},
  year   = {2016},
  doi    = {10.1021/acs.inorgchem.6b01398},
  note   = {PMID: 27726350},
  url    = {https://doi.org/10.1021/acs.inorgchem.6b01398},
  eprint = {https://doi.org/10.1021/acs.inorgchem.6b01398}
}

@article{dahlqvist2025,
author ="Dahlqvist, Martin and Rosen, Johanna",
title  ="Combined in- and out-of-plane chemical ordering in super-ordered MAX phases (s-MAX)",
journal  ="Nanoscale",
year  ="2025",
volume  ="17",
issue  ="22",
pages  ="13787-13796",
publisher  ="The Royal Society of Chemistry",
doi  ="10.1039/D5NR00672D",
url  ="http://dx.doi.org/10.1039/D5NR00672D"}

@article{bjork2023,
  author    = {Björk, Jonas and Halim, Joseph and Zhou, Jie and others},
  title     = {Predicting chemical exfoliation: fundamental insights into the synthesis of MXenes},
  journal   = {npj 2D Materials and Applications},
  volume    = {7},
  pages     = {5},
  year      = {2023},
  doi       = {10.1038/s41699-023-00370-8},
  url       = {https://doi.org/10.1038/s41699-023-00370-8}
}

@article{alameda2018,
author = {Alameda, Lucas
T. and Moradifar, Parivash and Metzger, Zachary P. and Alem, Nasim and Schaak, Raymond E.},
title = {Topochemical Deintercalation of Al from MoAlB: Stepwise Etching Pathway, Layered Intergrowth Structures, and Two-Dimensional MBene},
journal = {Journal of the American Chemical Society},
volume = {140},
number = {28},
pages = {8833-8840},
year = {2018},
doi = {10.1021/jacs.8b04705},
    note ={PMID: 29906120},

URL = { 
    
        https://doi.org/10.1021/jacs.8b04705
    
    

},
eprint = { 
    
        https://doi.org/10.1021/jacs.8b04705
    
    

}

}

@inbook{lambrinou2017,
author = {Lambrinou1, K. and Lapauw, T. and Tunca, B. and Vleugels, J.},
publisher = {John Wiley \& Sons, Ltd},
isbn = {9781119321811},
title = {MAX Phase Materials for Nuclear Applications},
booktitle = {Developments in Strategic Ceramic Materials II},
chapter = {21},
pages = {223-233},
doi = {https://doi.org/10.1002/9781119321811.ch21},
year = {2017}
}

@Article{chirica2021,
author ="Chirica, Iuliana M. and Mirea, Anca G. and Neaţu, Stefan and Florea, Mihaela and Barsoum, Michel W. and Neatu, Florentina",
title  ="Applications of MAX phases and MXenes as catalysts",
journal  ="J. Mater. Chem. A",
year  ="2021",
volume  ="9",
issue  ="35",
pages  ="19589-19612",
publisher  ="The Royal Society of Chemistry",
doi  ="10.1039/D1TA04097A",
url  ="http://dx.doi.org/10.1039/D1TA04097A"}

@article{li2022,
  author    = {Li, H. and Wang, Z. and Zou, N. and others},
  title     = {Deep-learning density functional theory Hamiltonian for efficient \emph{ab initio} electronic-structure calculation},
  journal   = {Nature Computational Science},
  volume    = {2},
  pages     = {367--377},
  year      = {2022},
  doi       = {10.1038/s43588-022-00265-6},
  url       = {https://doi.org/10.1038/s43588-022-00265-6}
}

@article{antunes2024,
  author    = {Antunes, L. M. and Butler, K. T. and Grau-Crespo, R.},
  title     = {Crystal structure generation with autoregressive large language modeling},
  journal   = {Nature Communications},
  volume    = {15},
  pages     = {10570},
  year      = {2024},
  doi       = {10.1038/s41467-024-54639-7},
  url       = {https://doi.org/10.1038/s41467-024-54639-7}
}

@article{zhao2023,
  title        = {Physics guided deep learning for generative design of crystal materials with symmetry constraints},
  author       = {Zhao, Y. and Siriwardane, E. M. D. and Wu, Z. and others},
  journal      = {npj Computational Materials},
  volume       = {9},
  pages        = {38},
  year         = {2023},
  publisher    = {Nature Publishing Group},
  doi          = {10.1038/s41524-023-00987-9},
  url          = {https://doi.org/10.1038/s41524-023-00987-9}
}

@misc{millerFlowMMGeneratingMaterials2024,
	title        = {{{FlowMM}}: {{Generating Materials}} with {{Riemannian Flow Matching}}},
	shorttitle   = {{FlowMM}},
	author       = {Miller, Benjamin Kurt and Chen, Ricky T. Q. and Sriram, Anuroop and Wood, Brandon M.},
	year         = 2024,
	month        = jun,
	publisher    = {arXiv},
	number       = {arXiv:2406.04713},
	doi          = {10.48550/arXiv.2406.04713},
	urldate      = {2025-09-04},
	eprint       = {2406.04713},
	primaryclass = {cs},
	archiveprefix = {arXiv}
}

@misc{pakornchoteDiffusionProbabilisticModels2023,
	title        = {Diffusion Probabilistic Models Enhance Variational Autoencoder for Crystal Structure Generative Modeling},
	author       = {Pakornchote, Teerachote and {Choomphon-anomakhun}, Natthaphon and Arrerut, Sorrjit and Atthapak, Chayanon and Khamkaeo, Sakarn and Chotibut, Thiparat and Bovornratanaraks, Thiti},
	year         = 2023,
	month        = aug,
	publisher    = {arXiv},
	number       = {arXiv:2308.02165},
	doi          = {10.48550/arXiv.2308.02165},
	urldate      = {2025-11-11},
	eprint       = {2308.02165},
	primaryclass = {cs},
	archiveprefix = {arXiv}
}

@misc{jiaoCrystalStructurePrediction2024,
	title        = {Crystal {{Structure Prediction}} by {{Joint Equivariant Diffusion}}},
	author       = {Jiao, Rui and Huang, Wenbing and Lin, Peijia and Han, Jiaqi and Chen, Pin and Lu, Yutong and Liu, Yang},
	year         = 2024,
	month        = mar,
	publisher    = {arXiv},
	number       = {arXiv:2309.04475},
	doi          = {10.48550/arXiv.2309.04475},
	urldate      = {2025-09-08},
	eprint       = {2309.04475},
	primaryclass = {cond-mat},
	archiveprefix = {arXiv}
}

@misc{jiaoSpaceGroupConstrained2024,
	title        = {Space {{Group Constrained Crystal Generation}}},
	author       = {Jiao, Rui and Huang, Wenbing and Liu, Yu and Zhao, Deli and Liu, Yang},
	year         = 2024,
	month        = apr,
	publisher    = {arXiv},
	number       = {arXiv:2402.03992},
	doi          = {10.48550/arXiv.2402.03992},
	urldate      = {2025-10-17},
	eprint       = {2402.03992},
	primaryclass = {cs},
	archiveprefix = {arXiv}
}

@article{klipfel_diff_aaai_2024,
	title        = {Vector Field Oriented Diffusion Model for Crystal Material Generation},
	author       = {Astrid Klipfel and Ya{\"{e}}l Fr{\'{e}}gier and Adlane Sayede and Zied Bouraoui},
	year         = 2024,
	journal      = {Proceedings of the AAAI Conference on Artificial Intelligence}
}

@misc{levySymmCDSymmetryPreservingCrystal2025,
	title        = {{{SymmCD}}: {{Symmetry-Preserving Crystal Generation}} with {{Diffusion Models}}},
	shorttitle   = {{SymmCD}},
	author       = {Levy, Daniel and Panigrahi, Siba Smarak and Kaba, S{\'e}kou-Oumar and Zhu, Qiang and Lee, Kin Long Kelvin and Galkin, Mikhail and Miret, Santiago and Ravanbakhsh, Siamak},
	year         = 2025,
	month        = may,
	publisher    = {arXiv},
	number       = {arXiv:2502.03638},
	doi          = {10.48550/arXiv.2502.03638},
	urldate      = {2025-11-11},
	eprint       = {2502.03638},
	primaryclass = {cond-mat},
	archiveprefix = {arXiv}
}

@misc{cornetKineticLangevinDiffusion2025,
	title        = {Kinetic {{Langevin Diffusion}} for {{Crystalline Materials Generation}}},
	author       = {Cornet, Fran{\c c}ois and Bergamin, Federico and Bhowmik, Arghya and Lastra, Juan Maria Garcia and Frellsen, Jes and Schmidt, Mikkel N.},
	year         = 2025,
	month        = jul,
	publisher    = {arXiv},
	number       = {arXiv:2507.03602},
	doi          = {10.48550/arXiv.2507.03602},
	urldate      = {2025-11-11},
	eprint       = {2507.03602},
	primaryclass = {cs},
	archiveprefix = {arXiv}
}

@article{chang2025space,
  title={Space Group Equivariant Crystal Diffusion},
  author={Chang, Rees and Pak, Angela and Guerra, Alex and Zhan, Ni and Richardson, Nick and Ertekin, Elif and Adams, Ryan P},
  journal={arXiv preprint arXiv:2505.10994},
  year={2025}
}

@inproceedings{nguyen2023hierarchical,
  title={Hierarchical gflownet for crystal structure generation},
  author={Nguyen, Tri Minh and Tawfik, Sherif Abdulkader and Tran, Truyen and Gupta, Sunil and Rana, Santu and Venkatesh, Svetha},
  booktitle={AI for Accelerated Materials Design-NeurIPS 2023 Workshop},
  year={2023}
}

@misc{parkExplorationCrystalChemical2024,
	title        = {Exploration of Crystal Chemical Space Using Text-Guided Generative Artificial Intelligence},
	author       = {Park, Hyunsoo and Onwuli, Anthony and Walsh, Aron},
	year         = 2024,
	month        = nov,
	doi          = {10.26434/chemrxiv-2024-rw8p5},
	urldate      = {2024-12-17},
	langid       = {english}
}

@article{ong2019accelerating,
  title={Accelerating materials science with high-throughput computations and machine learning},
  author={Ong, Shyue Ping},
  journal={Computational Materials Science},
  volume={161},
  pages={143--150},
  year={2019},
  publisher={Elsevier}
}

@article{zeni2025generative,
  title={A generative model for inorganic materials design},
  author={Zeni, Claudio and Pinsler, Robert and Z{\"u}gner, Daniel and Fowler, Andrew and Horton, Matthew and Fu, Xiang and Wang, Zilong and Shysheya, Aliaksandra and Crabb{\'e}, Jonathan and Ueda, Shoko and others},
  journal={Nature},
  volume={639},
  number={8055},
  pages={624--632},
  year={2025},
  publisher={Nature Publishing Group UK London}
}

@article{merchant2023scaling,
  title={Scaling deep learning for materials discovery},
  author={Merchant, Amil and Batzner, Simon and Schoenholz, Samuel S and Aykol, Muratahan and Cheon, Gowoon and Cubuk, Ekin Dogus},
  journal={Nature},
  volume={624},
  number={7990},
  pages={80--85},
  year={2023},
  publisher={Nature Publishing Group UK London}
}

@article{batatia2023,
  title={A foundation model for atomistic materials chemistry},
  author={Batatia, Ilyes and Benner, Philipp and Chiang, Yuan and Elena, Alin M and Kov{\'a}cs, D{\'a}vid P and Riebesell, Janosh and Advincula, Xavier R and Asta, Mark and Avaylon, Matthew and Baldwin, William J and others},
  journal={The Journal of chemical physics},
  volume={163},
  number={18},
  pages={184110},
  year={2025},
  publisher={AIP Publishing}
}

@article{batatia2025design,
  title={The design space of E (3)-equivariant atom-centred interatomic potentials},
  author={Batatia, Ilyes and Batzner, Simon and Kov{\'a}cs, D{\'a}vid P{\'e}ter and Musaelian, Albert and Simm, Gregor NC and Drautz, Ralf and Ortner, Christoph and Kozinsky, Boris and Cs{\'a}nyi, G{\'a}bor},
  journal={Nature Machine Intelligence},
  volume={7},
  number={1},
  pages={56--67},
  year={2025},
  publisher={Nature Publishing Group UK London}
}

@article{sanchez2018inverse,
	title        = {Inverse molecular design using machine learning: Generative models for matter engineering},
	author       = {Sanchez-Lengeling, Benjamin and Aspuru-Guzik, Al{\'a}n},
	year         = 2018,
	journal      = {Science},
	publisher    = {American Association for the Advancement of Science},
	volume       = 361,
	number       = 6400,
	pages        = {360--365}
}

@misc{mohantyCrysTextGenerativeAI2025,
	title        = {{{CrysText}}: {{A Generative AI Approach}} for {{Text-Conditioned Crystal Structure Generation}} Using {{LLM}}},
	shorttitle   = {{CrysText}},
	author       = {Mohanty, Trupti and Mehta, Maitrey and Sayeed, Hasan M. and Oded, Bat\_El and Pitussi, Itay and Borenstein, Arie and Srikumar, Vivek and Sparks, Taylor D.},
	year         = 2025,
	month        = nov,
	publisher    = {ChemRxiv},
	doi          = {10.26434/chemrxiv-2024-gjhpq-v2},
	urldate      = {2025-11-11},
	archiveprefix = {ChemRxiv},
	langid       = {english}
}

@misc{breuckGenerativeMaterialTransformer2025,
	title        = {A Generative Material Transformer Using {{Wyckoff}} Representation},
	author       = {Breuck, Pierre-Paul De and Piracha, Hashim A. and Rignanese, Gian-Marco and Marques, Miguel A. L.},
	year         = 2025,
	month        = jan,
	publisher    = {arXiv},
	number       = {arXiv:2501.16051},
	doi          = {10.48550/arXiv.2501.16051},
	urldate      = {2025-11-11},
	eprint       = {2501.16051},
	primaryclass = {cond-mat},
	archiveprefix = {arXiv}
}

@article{kazeev2503wyckoff,
	title        = {Wyckoff transformer: Generation of symmetric crystals, 2025},
	author       = {Kazeev, Nikita and Nong, Wei and Romanov, Ignat and Zhu, Ruiming and Ustyuzhanin, Andrey and Yamazaki, Shuya and Hippalgaonkar, Kedar},
	journal      = {URL https://arxiv. org/abs/2503.02407}
}

@article{alampara2024mattext,
	title        = {MatText: Do language models need more than text \& scale for materials modeling?},
	author       = {Alampara, Nawaf and Miret, Santiago and Jablonka, Kevin Maik},
	year         = 2024,
	journal      = {arXiv preprint arXiv:2406.17295}
}

@article{gruver2024fine,
  title={Fine-tuned language models generate stable inorganic materials as text},
  author={Gruver, Nate and Sriram, Anuroop and Madotto, Andrea and Wilson, Andrew Gordon and Zitnick, C Lawrence and Ulissi, Zachary},
  journal={arXiv preprint arXiv:2402.04379},
  year={2024}
}

@article{flam2023language,
  title={Language models can generate molecules, materials, and protein binding sites directly in three dimensions as xyz, cif, and pdb files},
  author={Flam-Shepherd, Daniel and Aspuru-Guzik, Al{\'a}n},
  journal={arXiv preprint arXiv:2305.05708},
  year={2023}
}

@misc{bone2025discoveryrecoverycrystallinematerials,
      title={Discovery and recovery of crystalline materials with property-conditioned transformers}, 
      author={Cyprien Bone and Matthew Walker and Kuangdai Leng and Luis M. Antunes and Ricardo Grau-Crespo and Amil Aligayev and Javier Dominguez and Keith T. Butler},
      year={2025},
      eprint={2511.21299},
      archivePrefix={arXiv},
      primaryClass={cond-mat.mtrl-sci},
      url={https://arxiv.org/abs/2511.21299}, 
}

@article{Leimeroth2025MSMSE,
  author  = {Leimeroth, Niklas and Erhard, Linus C. and Albe, Karsten and Rohrer, Jochen},
  title   = {Machine-learning interatomic potentials from a users perspective: a comparison of accuracy, speed and data efficiency},
  journal = {Modelling and Simulation in Materials Science and Engineering},
  year    = {2025},
  volume  = {33},
  number  = {6},
  pages   = {065012},
  doi     = {10.1088/1361-651X/adf56d},
  publisher = {IOP Publishing}
}

@misc{lemat_bulkunique_2024,
  author       = {Siron, Martin and
                  Djafar, Inel and
                  Ritchie, Lucile and
                  Du-Fayet, Etienne and
                  Rossello, Amandine and
                  Ramlaoui, Ali and
                  von Werra, Leandro and
                  Wolf, Thomas and
                  Duval, Alexandre},
  title        = {{Lemat-bulkunique dataset}},
  year         = {2024},
  note         = {Dataset}
}

\end{document}